# Optimal Resource Allocation in Distributed Broadband Wireless Communication Systems

Yao Yao, Mustafa Mehmet-Ali, Shahin Vakilinia

Department of Electrical & Computer Engineering, Concordia University
Montreal, Canada
E-mail: {y_yao11,s_vakili}@encs.concordia.ca, mustafa@ece.concordia.ca,

*Abstract*—This paper is concerned with optimization of distributed broadband wireless communication (BWC) systems. BWC systems contain a distributed antenna system (DAS) connected to a base station with optical fiber. Distributed BWC systems have been proposed as a solution to the power constraint problem in traditional cellular networks. So far, the research on BWC systems have advanced on two separate tracks, design of the system to meet the quality of service requirements (QoS) and optimization of location of the DAS. In this paper, we consider a combined optimization of BWC systems. We consider uplink communications in distributed BWC systems with multiple levels of priority traffic with arrivals and departures forming renewal processes. We develop an analysis that determines packet delay violation probability for each priority level as a function of the outage probability of the DAS through the application of results from renewal theory. Then, we determine the optimal locations of the antennas that minimize the antenna outage probability. We also study the trade off between the packet delay violation probability and packet loss probability. This work will be helpful in the designing of the distributed BWC systems.

## I. INTRODUCTION

The future services in wireless networks will be requiring higher transmission rates. This rate is expected to rise to 1Gbits/s for the next generation communication systems according to the objective of IMT-Advanced (International Mobile Telecommunications) system, which is set by ITU-R (International Telecommunication Union-Radiocommunication Standardization Sector) [1-3]. However, the constraint on the transmit power limits the transmission rate, especially in the uplink communications from user to base station. One promising solution to this problem is the distributed broadband wireless communications (BWC) systems [2].

A distributed BWC system is made up of a distributed antenna system (DAS) and the radio over fiber (RoF) technology [2]. In DAS, antennas are placed at geographically separate locations in the cell [2], which is a contrary to the so called centralized antenna system (CAS) that having an antenna set at the center of a cell. The RoF technology, which is very reliable and has very small delay, is responsible for providing communication between these antennas and a central processor where they are jointly processed [2]. In this way, a user is very likely to have an antenna nearby compared to CAS, which will in turn reduce the transmission power requirement of that user. What's more, it's possible to place many distributed antennas in each cell due to the relatively low cost of distributed antennas [4].

By adopting DAS, a BWC system can solve the power constraint problem naturally instead of shrinking the cell size as in conventional CAS, which results in larger overhead and delay due to higher frequency of handovers between cells.

There has been a number of research works on distributed BWC systems. These work have dealt with two main problems, which are determining the antenna set that meet QoS metrics such as delay and packet loss and the other optimal placement of the antennas.

[5] considers the downlink transmission, from base station to the user, and assumes that the central server at the base station maintains a separate queue for each of the users. This work considers selection of a subset of the available distributed antennas such that the probability of packet queuing delay is less than a threshold. [6] considers uplink transmission in a distributed BWC system. Again, the objective has been selection of a subset of distributed antennas at each user to keep packet loss probability below a threshold value. In both [5] and [6], locations of the antennas are fixed and the objective function has not been optimized with respect to (wrt) the location of the antennas.

Optimal placement of antennas has been considered in other works but independent of QoS metrics. In [4, 7], the objective has been to optimize the sum-rate uplink capacity of a single cell with multiple users. It is assumed that the antennas are placed on a circle centered at the cell center and their locations on the circle are uniformly distributed independent of each other. The optimal radius of the antenna circle has been determined, however, this work is more appropriate to the CDMA systems since all the users in the cell will be transmitting simultaneously. Further, the paper considers the effects of inter-cell interference only by simulation.

In [8], the optimal location of antennas in a cellular setting has been studied for the downlink transmission based on the performance metrics of capacity and power efficiency. This work takes the inter-cell interference into account. Though no closed form results are obtained, their numerical results show that for some certain cases, the optimal layout consists of placement of one antenna at center and remaining ones on a circle around it, and that circle shrinks as the interference coefficient increases. This work does not provide any results on the positions of the antennas on the circle.

In this paper, we consider the uplink communications and assume that each user has multiple levels of priority traffic with different QoS requirements. We determine the packet delay violation probability through the use of results from renewal theory for each type of traffic as a function of outage probability. Then, we determine the optimal location of antennas that minimizes the antenna outage probability. We note that our work is more general compared to [5, 6], because it considers multiple types of traffic with different QoS requirements. In relation to optimal placement of antennas, our work is closer to that in [8], but we consider the uplink as opposed to downlink communications, which is more critical due to limited transmit power of the users.

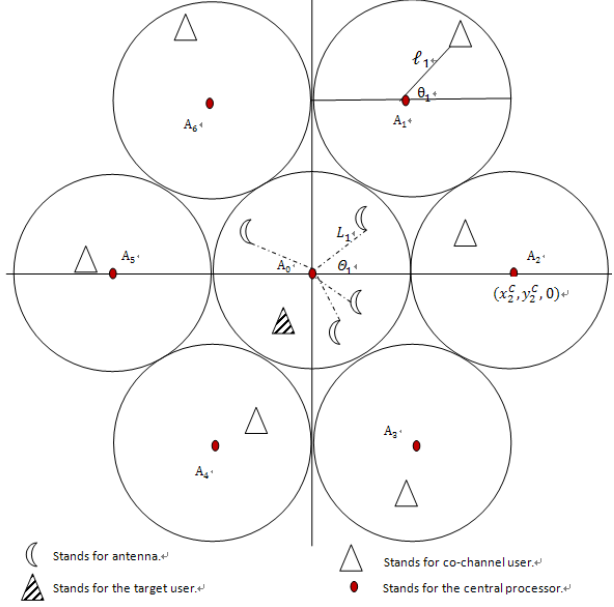

Fig. 1. Topology of the distributed BWC system which has a cellular network architecture

The outline of the remainder of the paper is as follows. Section Ⅱ. describes our system model. Section Ⅲ presents the queuing delay analysis for multiple types of traffic flows. Section Ⅳ gives the application of these results to a system with two types of traffic flows. Section V. extends the application to a system with four types of traffic flows. Next, section VI. presents the work on the optimal placement of the antennas. Section VII presents the numerical and simulation results regarding the analysis. And finally section VIII gives the conclusions of the paper.

## II. SYSTEM MODEL

### A. Network Topology

We consider a cellular network architecture with distributed antenna system. Each cell together with its neighbors are referred to as a cluster. The cells in a cluster will be numbered as $A_j$, $0 \leq j \leq F\text{-}1$, where $F$ is the cluster size. $A_0$ will denote the cell at the center of the cluster and it will be referred as the target cell. Fig. 1 shows an example network with a cluster of seven cells. Each cell contains a central processor and $M$ distributed antennas, which are placed at different geographical locations and are linked to the central processor through RoF technology. In the figure only the antennas in the target cell $A_0$ have been shown to prevent crowding, however, each of the other cells also contains $M$ distributed antennas.

It will be assumed that all the antennas have the same height $\hbar$, and the users are all at the ground level. In the target cell, locations of the $M$ antennas wrt the cell center will be denoted by $(L_m, \Theta_m, \hbar)$ in polar coordinates, for $1 \leq m \leq M$. We number the antennas according to their increasing polar angles, thus, $\theta_m > \theta_{m-1}$. Define antennas location vector $\mathbf{L} = ((L_1, \Theta_1, \hbar), (L_2, \Theta_2, \hbar), \dots (L_M, \Theta_M, \hbar))$

### B. Frequency Reuse Model

We will assume that all the frequencies will be available in every cell, thus the reuse factor will be one as in [8]. The available spectrum will be divided into a number of channels and each user will be assigned a single channel at any time. We will also assume that the system is saturated, so that all the channels will always be busy, and the users using the same channel in the neighboring cells will be interfering with each other. This corresponds to the worst case scenario of co-channel interference which is the major cause of an outage in a cellular system [9]. From Fig. 1, a user in the target cell will be considered as a target user and the users in the neighboring cells using the same channel are considered as the interferers.

Thus there will be a single user in each cell using the same channel in a cluster. Let user $i$ refer to the user in cell $A_i$, who is using a particular channel in the cluster. We will let $(\ell_i, \theta_i, 0)$ denote the location of user $i$ in polar coordinates relative to the center of its home cell and $(x_i, y_i, 0)$ denote its location in Cartesian coordinates relative to the center of cell $A_0$. Then, the latter may be expressed in terms of the former as follows,

$$(x_i, y_i, 0) = (x_i^C + \ell_i \cos \theta_i, y_i^C + \ell_i \sin \theta_i, 0) \text{ for } 0 \leq i \leq F\text{-}1.$$

### C. Transmission Model

The time-axis is slotted and the transmission of a packet always starts at the beginning of a slot and it takes one slot. A user will broadcast its packet which will be received by the antennas in its cell, the received signals will be locally decoded and the results will be forwarded to the central processor through fiber lines, respectively. If at least one of the antennas can decode the packet successfully, the central processor will receive the transmitted packet correctly. On the other hand, if none of the antennas is able to decode the packet successfully, then the central processor will drop this packet or ask for retransmissions, depending on the type of the packet.

## III. QUEUING DELAY

In this section, we will determine the queuing delay bounds for a user with multiple types of traffic with different QoS requirements. We will assume that a user has $N$ types of traffic flows, numbered as 1≤n≤N. The packets of flows are stored in separate queues. Each traffic flow is assigned a priority level, which increases with decreasing flow number. A packet of a traffic flow is only served if the queues of higher priority traffic flows are empty. Let $x_n$, $y_n$ denote independent and identically distributed inter-arrival time and service time of the packets of traffic flow $n$ respectively, if the system was serving only this traffic flow. Let $\mu_{x_n}$, $\sigma^2_{x_n}$ and $\mu_{y_n}$, $\sigma^2_{y_n}$ denote the mean and variance of the random variables $x_n$, $y_n$ respectively. We note that the service time of a packet begins when it reaches to the head of its queue and completed either following its successful transmission or discarding by the system.

In this section, we will determine the queuing delay bounds of packets of different flows. We will use the delay bound violation probability from the theory of effective bandwidth in our analysis [10, 11]. The same metric had also been used in [5]. The delay-bound violation probability is defined as the probability that the packet delay exceeds a certain value

$$Prob(D_n > D_{th}^n) \quad (1)$$

for the traffic flow $n$, where $D_n$ and $D_{th}^n$ represent the queuing delay and a queuing delay threshold respectively. The delay bound is determined through the utilization of energy function, which is defined as the asymptotic log moment generating function of arrival and service processes [10].

Initially, we will assume that the system is operating with a single type of traffic, $n^{th}$ traffic flow, and later on the results will be extended to multiple types of traffic. Let $A_n(t)$ denote the number of packet arrivals to the queue during the time interval [0, t). Defining $C_n(t)$ as the number of departures from the queue during the interval [0, t) assuming that it is under saturation. Let $\Lambda_n(\varphi)$ and $\Gamma_n(\varphi)$, $(\varphi > 0)$, denote the energy functions of the arrival and service processes [10],

$$\Lambda_n(\varphi) = lim_{t \to \infty} \frac{1}{t} \log E(e^{\varphi A_n(t)}) \quad (2)$$
$$\Gamma_n(\varphi) = lim_{t \to \infty} \frac{1}{t} \log E(e^{\varphi C_n(t)}) \quad (3)$$

From [5], we have the delay-bound violation probability for the packets of $n^{th}$ traffic flow,

$$Prob(D_n > D_{th}^n) \approx e^{-\Lambda_n(\varphi^*)D_{th}^n} \quad (4)$$

where $\varphi^*$ is the unique solution of $\Lambda_n(\varphi) + \Gamma_n(-\varphi) = 0$.

In general, it may not be possible to find the energy functions defined in (2-3), which limits the applicability of this method. In that case, we propose to use the asymptotic central limit theorem for renewal processes in their evaluations, which states that in a renewal process, the number of renewal points asymptotically approaches to a Gaussian distribution with mean $\frac{t}{\mu}$, and variance $\frac{\sigma^2 t}{\mu^3}$ as $t$ goes to infinity where $\mu$ and $\sigma^2$ are the mean and variance of the renewal interval [12]. The application of this result gives the energy function of the arrival and service processes as follows,

$$\Lambda_n(\varphi) = \frac{\varphi}{\mu_{x_n}} + \frac{\varphi^2 \sigma^2_{x_n}}{2\mu^3_{x_n}}, \quad \Gamma_n(\varphi) = \frac{\varphi}{\mu_{y_n}} + \frac{\varphi^2 \sigma^2_{y_n}}{2\mu^3_{y_n}} \quad (5)$$

Next, we will extend the results to a user with multiple types of traffic. In this scenario, the $n^{th}$ traffic flow will have priority level $n$. The arrival process of this traffic flow will remain as before but its service process will change because of the services given to the packets of higher priority flows. Let $\tilde{C}_n(t)$ denote the number of departures from the $n^{th}$ flow during the interval [0, t),

$$\tilde{C}_n(t) = C_n(t - \sum_{j=1}^{n-1} Y_j(t)) \quad (6)$$
where, $\quad Y_j(t) = \sum_{i=1}^{A_j(t)} y_{j,i}$

with $y_{j,i}$ denoting the service time of the $i^{th}$ packet from the $j^{th}$ flow. Then the above may be written as,

$$\tilde{C}_n(t) = C_n(t) - \sum_{j=1}^{n-1} C_n(Y_j(t)) \quad (7)$$

Next letting $\tilde{\Gamma}_n(\varphi)$ denote the energy function of the service process $\tilde{C}_n(t)$,

$$\tilde{\Gamma}_n(\varphi) = lim_{t \to \infty} \frac{1}{t} \log E\left[e^{\varphi C_n(t) - \varphi \sum_{j=1}^{n-1} C_n(Y_j(t))}\right] \quad (8)$$

We will assume that the system is stable, thus none of the queues are starving for service. Then, we have $t > \sum_{j=1}^{n-1} Y_j(t)$, as a result, the random variables $Y_j(t)$ with different $j$ will be independent of each other. Thus,

$$\tilde{\Gamma}_n(\varphi) = lim_{t \to \infty} \frac{1}{t} \log \left\{ E[e^{\varphi C_n(t)}] \prod_{j=1}^{n-1} E\left[e^{-\varphi C_n(Y_j(t))}\right] \right\} \quad (9)$$

Expressing the above as sum of two logarithms and substituting from (3),

$$\tilde{\Gamma}_n(\varphi) = \Gamma_n(\varphi) + \lim_{t \to \infty} \frac{1}{t} \left\{ \sum_{j=1}^{n-1} \log E\left[ E[e^{-\varphi C_n(k)} | Y_j(t)] \right] \right\} \quad (10)$$

Again, application of the central limit theorem for the renewal processes to the conditional expectation gives,

$$E\left[E[e^{-\varphi C_n(k)}|Y_j(t)]\right] = E\left[e^{Y_j(t)\left(-\frac{\varphi}{\mu_{y_n}} + \frac{\varphi^2 \sigma^2_{y_n}}{2\mu^3_{y_n}}\right)}\right] \quad (11)$$

From its definition in (6), $Y_j(t)$ is a random sum with mean and variance given by,

$$\mu_{Y_j(t)} = \frac{\mu_{y_j} t}{\mu_{x_j}}, \quad \sigma^2_{Y_j(t)} = \mu^2_{y_j} \frac{\sigma^2_{x_j} t}{\mu^3_{x_j}} + \frac{\sigma^2_{y_j} t}{\mu_{x_j}} \quad (12)$$

From the general central limit theorem, $Y_j(t)$ approaches to a normal distribution for large values of $t$ with mean and variance given in the above. Then, the expectation on the RHS of (11) corresponds to the moment generating function of a normal random variable, as a result (10) becomes,

$$\tilde{\Gamma}_n(\varphi) = \Gamma_n(\varphi) + \lim_{t\to\infty}\frac{1}{t}\left\{\sum_{j=1}^{n-1} \log e^{\phi\mu_{Y_j(t)}+\frac{\phi^2\sigma^2_{Y_j(t)}}{2}}\Big|_{\phi=-\frac{\varphi}{\mu_{y_n}}+\frac{\varphi^2\sigma^2_{y_n}}{2\mu^3_{y_n}}}\right\}$$

Substituting from (5,12) in the above,

$$\tilde{\Gamma}_n(\varphi) = \frac{\varphi}{\mu_{y_n}} + \frac{\varphi^2\sigma^2_{y_n}}{2\mu^3_{y_n}} + \left\{\Sigma_{j=1}^{n-1}\phi\frac{\mu_{y_j}}{\mu_{x_j}} + \frac{\phi^2}{2}\left(\mu^2_{y_j}\frac{\sigma^2_{x_j}}{\mu^3_{x_j}} + \frac{\sigma^2_{y_j}}{\mu_{x_j}}\right)\Big|_{\phi=-\frac{\varphi}{\mu_{y_n}}+\frac{\varphi^2\sigma^2_{y_n}}{2\mu^3_{y_n}}}\right\} \quad (13)$$

This completes the derivation of the energy function of the service process of the $n^{th}$ flow.

Next, we will demonstrate the accuracy of the asymptotic central limit theorem for renewal processes through an example. Let us consider the Binomial process, whose exact energy function is known [11]. The exact and asymptotic energy functions of a Binomial process are given by,

$$\Gamma(\varphi) = \log(q + (1-q)e^\varphi), \quad \Gamma^*(\varphi) = (1-q)\varphi\left(1+\frac{1}{2}q\varphi\right)$$

where $1-q$ is the success probability. These two results have been plotted in Fig. 2 as a function of $\varphi$. It may be seen that they are very close to each other, which gives confidence to the use of asymptotic energy function.

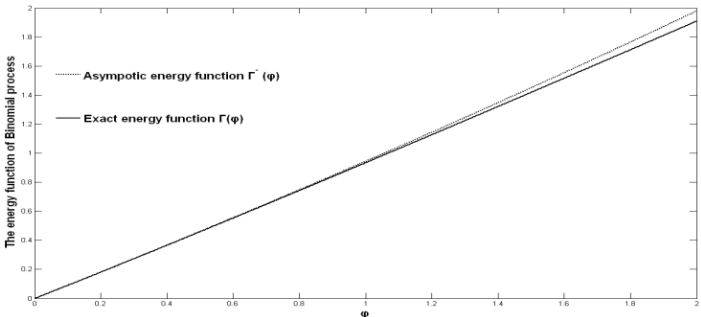

Fig.2. Comparison of asymptotic and exact energy functions of Binomial process with success probability of $1-q$ for $q=0.1$

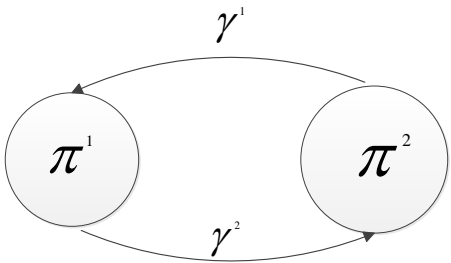

Fig. 3: A two-state Markov fluid. $\gamma^i$ is the transition rate from state $i$ to state the other state. $\pi^i$ is the probability that the system will be at state $i$.

## IV. A SYSTEM WITH TWO TYPES OF TRAFFIC FLOWS

Next, we will apply the above results to a system with two types of traffic flows, which will be referred as voice and data respectively. The voice and data packets will be stored in separate queues and voice will be given higher priority wrt data traffic. It will be assumed that a transmitted voice packet that can not be decoded will be dropped since voice is delay sensitive. On the other hand, the failed data packets will be retransmitted and will be lost only after $L$ unsuccessful transmissions. In the following, we will study the delay and packet losses of data flow since voice flow is redundant and is given higher priority. Assume that both voice and data arrive according to independent Poisson processes with rates $\lambda_V$ and $\lambda_D$ packets/sec respectively. The exact energy function of a Poisson process is known, [11], giving the energy functions of voice and data as,

$$\Lambda_1(\varphi) = \lambda_V(e^\varphi - 1), \quad \Lambda_2(\varphi) = \lambda_D(e^\varphi - 1)$$

From (3) and (13), the energy functions of the voice and data departure processes are respectively,

$$\Gamma_1(\varphi) = \tilde{\Gamma}_1(\varphi) = \varphi; \quad \tilde{\Gamma}_2(\varphi) = \frac{\varphi}{\mu_{y_2}} + \frac{\varphi^2\sigma^2_{y_2}}{2\mu^3_{y_2}} + \lambda_V\left(e^{\frac{-\varphi}{\mu_{y_2}}+\frac{\varphi^2\sigma^2_{y_2}}{2\mu^3_{y_2}}} - 1\right)$$

Next, we determine $\mu_{y_2}, \sigma^2_{y_2}$ which are the mean and variance of the service time of data packets. The service time of the data packets has a truncated geometric distribution,

$$\begin{cases} Prob(y_2 = i \text{ slots}) = (1-p)p^{i-1} \text{ for } i < L \\ Prob(y_2 = i \text{ slots}) = p^{L-1} \text{ for } i = L \end{cases}$$

where $p$ is the probability that transmission of a packet will result in an outage. Let $M_D(z)$ denote the PGF of the above service time distribution,

$$M_D(z) = E[z^{y_2}] = (1-p)z\frac{1-(zp)^{L-1}}{1-zp} + z^L p^{L-1} \quad (14)$$

From the above PGF,

$$\mu_{y_2} = \frac{1-p^L}{1-p}, \quad \sigma^2_{y_2} = \frac{-p^{2L}-p^{L+1}(1+2L)+p^L(1-2L)+2p}{(1-p)^2}$$

Finally, the delay violation probability of data packets is given by,

$$Prob(D_2 > D^2_{th}) \approx e^{-\Lambda_2(\varphi^*)D^2_{th}}$$

where $\varphi^*$ is the unique solution of the equation $\Lambda_2(\varphi) + \tilde{\Gamma}_2(-\varphi) = 0$. The packet loss probability of data packets is given by,

$$Prob(packet\ loss) = p^L$$

## V. A SYSTEM WITH FOUR TYPES OF TRAFFIC FLOWS

In this section, we will extend the analysis to a system with four types of traffic flows by adding a multimedia A and a multimedia B traffic flow into the formal system model.

In the new system, multimedia A is given a lower priority than the voice traffic but higher than the multimedia B, while multimedia B is given a higher priority than the data traffic. We assume multimedia A is a two-state Markov fluid which is characterized by four parameters: $\lambda^1_A$, $\lambda^2_A$, $\gamma^1_A$ and

$\gamma_A^2$. Where $\lambda_A^1$, $\lambda_A^2$ refers to the Poisson packet arrival rate at state 1 and 2, and $\gamma_A^1, \gamma_A^2$ refers to the transition rate from state 1 to 2 and from 2 to 1 respectively. Similar thing is for the multimedia B traffic with suffix B. We also assume that service of multimedia packets will be the same as the service of the voice packets.

Let $g_2(t)$ denote the pdf of the inter-arrival time of the packets of the 2nd traffic flow, the multimedia A, then, it is given by,

$$g_{x_2}(t) = \pi_A^1 \lambda_A^1 e^{-\lambda_A^1 t} + \pi_A^2 \lambda_A^2 e^{-\lambda_A^2 t} \quad (15)$$

where $\pi_A^1, \pi_A^2$ are the steady-state probabilities of the two-state Markov source being in states 1 and 2 respectively. These probabilities are given by $\pi_A^1 = \frac{\gamma_A^2}{\gamma_A^1 + \gamma_A^2}$, $\pi_A^2 = 1 - \pi_A^1$. Next we can determine the mean and variance of $x_2$ by obtaining the first and second moment of the Laplace transform of the above expression. Then we have,

$$\mu_{x_2} = \frac{\pi_A^1}{\lambda_A^1} + \frac{\pi_A^2}{\lambda_A^2}$$

$$\sigma_{x_2}^2 = \left(\frac{1}{\lambda_A^1}\right)^2 \left(2\pi_A^1 - \pi_A^{1^2}\right) + \left(\frac{1}{\lambda_A^2}\right)^2 \left(2\pi_A^2 - \pi_A^{2^2}\right)$$

Same thing is for the 3rd traffic flow, the multimedia B. And thus we can have the following equation set:

$$\begin{cases} \mu_{x_1} = \frac{1}{\lambda_V}, \quad \mu_{y_1} = 1 \\ \sigma_{x_1}^2 = \frac{1}{\lambda_V^2}, \sigma_{y_1}^2 = 0 \\ \mu_{x_2} = \frac{\pi_A^1}{\lambda_A^1} + \frac{\pi_A^2}{\lambda_A^2}, \quad \mu_{y_2} = 1 \\ \sigma_{x_2}^2 = \left(\frac{1}{\lambda_A^1}\right)^2 \left(2\pi_A^1 - \pi_A^{1^2}\right) + \left(\frac{1}{\lambda_A^2}\right)^2 \left(2\pi_A^2 - \pi_A^{2^2}\right), \sigma_{y_2}^2 = 0 \\ \mu_{x_3} = \frac{\pi_B^1}{\lambda_B^1} + \frac{\pi_B^2}{\lambda_B^2}, \quad \mu_{y_3} = 1 \\ \sigma_{x_3}^2 = \left(\frac{1}{\lambda_B^1}\right)^2 \left(2\pi_B^1 - \pi_B^{1^2}\right) + \left(\frac{1}{\lambda_B^2}\right)^2 \left(2\pi_B^2 - \pi_B^{2^2}\right), \sigma_{y_3}^2 = 0 \\ \mu_{x_4} = \frac{1}{\lambda_D}, \quad \mu_{y_4} = \frac{1-p^L}{1-p} \\ \sigma_{x_4}^2 = \frac{1}{\lambda_D^2}, \sigma_{y_4}^2 = \frac{-p^{2L} - p^{L+1}(1+2L) + p^L(1-2L) + 2p}{(1-p)^2} \end{cases}$$

(16)

Submitting (16) into (13), we could have $\tilde{\Gamma}_4(\varphi)$. And finally the delay violation probability of data packets is given by,

$$Prob(D_4 > D_{th}^4) \approx e^{-\Lambda_4(\varphi^*)D_{th}^4}$$

where $\varphi^*$ is the unique solution of the equation

$$\Lambda_4(\varphi) + \tilde{\Gamma}_4(-\varphi) = 0.$$

## VI. OPTIMAL DISTRIBUTED ANTENNA PLACEMENT

In this section, we will determine optimal locations of antennas by minimizing system's outage probability seen by a target user. First, we will determine the signal to noise ratio of the target user, and then we will derive the probability of system outage.

The distance between user $i$ and the $m^{th}$ antenna is given by,

$$\rho_{m,i} = \sqrt{(x_i - L_m cos(\Theta_m))^2 + (y_i - L_m sin(\Theta_m))^2 + \hbar^2} \quad (17)$$

$0 \leq i \leq F - 1$ and $1 \leq m \leq M$.

In the above, $2\lambda$ is the path loss exponent. $h_{m,i}$ are independent identically distributed (i.i.d) complex Gaussian random variables of Rayleigh fading channels. Since $|h_{m,i}|$ has a Rayleigh distribution, $|h_{m,i}|^2$ has an exponential distribution with mean equal to one.

In the following, we treat interferences from other co-channel users as noises, which are assumed to be the only source of noise. We consider that ach channel is time slotted such that the neighbor cell users communicate with in a synchronous manner. Meanwhile, a user, in a main cell is imposed by interference of the neighbor-cells users. In this paper, we mainly focus on the scenarios where user utilizes the channel used by set of neighbor-cells users. This implies that channel interference level sensed by the user is a random variable dependent on the other users' transmission states. We model each channel as an ON-OFF source alternating between state ON (active) and state OFF (inactive). An ON/OFF channel usage model specifies a time slot in which the neighbor user signals is or isn't transmitting over the channel. When more neighbor users are in the OFF time slot, main user transmission is affected less interference. Suppose that each channel changes its state independently. Let $\alpha_i$ be the probability that the $i_{th}$ neighbor user channel exists in state ON, Then, the channel state can be characterized by a two-state Markov chain.

Let $W$ denote the received signal strength at an antenna when a user is transmitting at a distance 1 from that antenna. Then $S_{m,i}$, the received signal strength at the antenna $m$ from user $i$, can be written as,

$$S_{m,i} = \begin{cases} W(r_{m,i})^{-2\lambda} |h_{m,i}|^2 & \text{with probability } a, 0 \leq i \leq F-1 \\ 0 & \text{with probability } 1-a \end{cases}$$

(18)

As a result the instantaneous SNR, $\Gamma_m$, at the antenna m is given by,

$$\Gamma_m = \frac{S_{m,0}}{\sum_{i=1}^{F-1} S_{m,i}} = \frac{\mathbb{X}_{m,0}}{\mathbb{Y}_m} \quad (19)$$

where, $\mathbb{X}_{m,i} = \frac{S_{m,i}}{W}$, $\mathbb{Y}_m = \sum_{i=1}^{F-1} \mathbb{X}_{m,i}$

In general, the outage probability at the antenna $m$ can be expressed as [13],

$$Prob_m(outage) = Prob(I_m < R) \quad (20)$$

where $I_m$ denotes the mutual information between the transmitter and receiver antenna $m$, and $R$ is the required transmission rate or spectrum efficiency in bits/sec/Hz [6]. From [14], $I_m$ can be expressed as,

$$I_m = log(1 + \Gamma_m) \quad (21)$$

Substituting (21) in the (20),

$$Prob_m(outage) = P(\Gamma_m < 2^R - 1) \quad (22)$$

Next let us define,

$$K = 2^R - 1, \quad \mathbb{Z}_m = K\mathbb{Y}_m - \mathbb{X}_{m,0}$$

Then, substituting (19) in (22), $Prob_m(outage)$ may be expressed as,

$$Prob_m(outage) = Prob[\mathbb{Z}_m > 0] \quad (23)$$

We continue our analysis by setting $\alpha = 1$. For Let $f_{\mathbb{Z}_m}(x)$ denote the pdf of the random variable (rv) $\mathbb{Z}_m$ and $\mathbb{Z}_m(s)$ its Laplace transform,

$$\mathbb{Z}_m(s) = \mathbb{Y}_m(Ks)\mathbb{X}_{m,0}(-s) = \mathbb{Y}_m(Ks)\mathbb{X}_{m,0}(-s) \quad (24)$$

where $\mathbb{Y}_m(s)$ and $\mathbb{X}_{m,0}(s)$ are Laplace transform of the rvs $\mathbb{Y}_m$ and $\mathbb{X}_{m,0}$. From (19), $\mathbb{Y}_m$ is the sum of $F$-1 independent exponentially distributed rvs, we have,

$$\mathbb{Y}_m(s) = \prod_{i=1}^{F-1} \mathbb{X}_{m,i}(s) \quad (25)$$

where $\mathbb{X}_{m,i}(s) = E[e^{-s\mathbb{X}_{m,i}}] = \frac{(\rho_{m,i})^{2\lambda}}{s + (\rho_{m,i})^{2\lambda}} \alpha + 1 - \alpha$, for $0 \le i \le F - 1$

Substituting (25) in (24) gives,

$$\mathbb{Z}_m(s) = \frac{(\rho_{m,0})^{2\lambda}}{-s + (\rho_{m,0})^{2\lambda}} \prod_{i=1}^{F-1} \frac{(\rho_{m,i})^{2\lambda}}{Ks + (\rho_{m,i})^{2\lambda}} = H_m \mathcal{F}_m(s) \quad (26)$$

where,

$$H_m = -\prod_{i=0}^{F-1}(\rho_{m,i})^{2\lambda}(K)^{-F+1}, \quad \mathcal{F}_m(s) = \prod_{n=1}^{N} \frac{1}{(s+c_{m,n})^{k_n}} \quad (27)$$

with $c_{m,n}$, $k_n$ denoting a distinct value and its frequency in the set $\{-(\rho_{m,0})^{2\lambda}, \frac{1}{K}(\rho_{m,1})^{2\lambda} \dots, \frac{1}{K}(\rho_{m,F-1})^{2\lambda}\}$ where $1 \le n \le N$ with $1 \le N \le F$, and $c_{m,1} = -(\rho_{m,0})^{2\lambda}$.

Now, expanding the above in a partial fraction,

$$\mathbb{Z}_m(s) = H_m \sum_{n=1}^{N} \sum_{j=1}^{k_n} \frac{b_{m,n}^j}{(s+c_{m,n})^j} \quad (28)$$

where, $b_{m,n}^j = \frac{1}{(j-1)!} \frac{d^{j-1}[(s+c_{m,n})^j \mathcal{F}_m(s)]}{ds^{j-1}} \bigg|(s = -c_{m,n})$

A Taking the inverse transform of the above gives the pdf of the rv $\mathbb{Z}_m$,

$$f_{\mathbb{Z}_m}(x) = H_m \left\{ b_{m,1}^1 e^{-c_{m,1}x} u[-x] + \sum_{n=2}^{N} \sum_{j=1}^{k_n} \frac{b_{m,n}^j}{(j-1)!} x^{j-1} e^{-c_{m,n}x} u[x] \right\}$$

where $u[x] = \begin{cases} 1, & x \ge 0 \\ 0, & x < 0 \end{cases}$

From (23), the probability of outage of antenna $m$ is given by,

$$Prob_m(outage) = \int_0^{+\infty} f_{\mathbb{Z}_m}(x)dx = H_m \sum_{n=2}^{N} \sum_{j=1}^{k_n} \frac{b_{m,n}^j}{(c_{m,n})^j} \quad (29)$$

We assume that outages of the antennas are independent of each other, therefore, the system outage probability, $\mathbb{P}(outage)$, for a fixed set of user locations is given by,,

$$\mathbb{P}(outage) = \prod_{m=1}^{M} Prob_m(outage) \quad (30)$$

Next, unconditioning the above result wrt the location vector of the users $(\boldsymbol{\ell}, \boldsymbol{\theta}, 0) = \{(\ell_0, \theta_0, 0), (\ell_1, \theta_1, 0), \dots, (\ell_{F-1}, \theta_{F-1}, 0)\}$, we have,

$$E(\mathbb{P}) = \int_{\ell_0=0}^{1} \dots \int_{\ell_{F-1}=0}^{1} \int_{\theta_0=0}^{2\pi} \dots \int_{\theta_{F-1}=0}^{2\pi} \mathbb{P}(outage) \prod_{i=0}^{F-1} p(\ell_i) p(\theta_i) d\ell_i d\theta_i \quad (31)$$

where $p(\ell_i), p(\theta_i)$ are the marginal pdfs of the coordinates of user $i$, $\ell_i$ and $\theta_i$ respectively. Since we are assuming that user locations are uniformly distributed in a cell we have, $Prob(\ell < \ell_i) = \frac{\ell_i^2}{r^2}$, and taking into account that $r$=1, we have,

$$p(\ell_i) = 2\ell_i, \text{for } 0 \le \ell \le 1, \quad p(\theta_i) = \frac{1}{2\pi}, \text{for } 0 \le \theta < 2\pi \quad (32)$$

We define the antenna locations that minimize the expected outage probability of the system in (31) as being optimal. Unfortunately, it has not been possible to obtain the minimum of the expected outage probability of the system analytically due to complexity of this function. As a result, we have determined minimum value of the expected outage probability of the system numerically by applying the Robbins-Monro stochastic approximation method as in [8].

## VII. NUMERICAL RESULTS

In this section, we present some numerical results regarding the analysis in the paper and simulation results to verify the accuracy of the analysis.

First, we present numerical and simulation results regarding the a system with two types of traffics which consist a voice traffic and a data traffic. It will be assumed that a transmitted voice packet that can not be decoded will be dropped since voice is delay sensitive. On the other hand, the failed data packets will be retransmitted and a data packet will be dropped only after $L$ unsuccessful transmissions. Dropped packets will be treated as packet loss. Both voice and data arrive according to independent Poisson processes with rates $\lambda_V$ and $\lambda_D$ packets/sec respectively.

Fig.s 3, 4 present the delay violation probabilities of data packets as a function of the delay threshold for the data traffic with $\lambda_V$, $\lambda_D$, p and L as parameters. From Fig. 3. it may be seen that increasing $\lambda_V$ increases the delay violation probability for a fixed value of $\lambda_D$. Similarly, from Fig. 4, increasing $\lambda_D$ increases the delay violation probability for a fixed value of $\lambda_V$.

In Fig 5, we plot numerical and simulation results of delay violation probability to show the accuracy of our analysis. As may be seen, there is a very good match between numerical and simulation results. This increases the confidence in the validity of the analysis, specially, in the use of asymptotic results from the renewal theory.

Fig. 6 presents the delay violation probabilities of data packets as a function of the data delay threshold with total number of transmissions, $L$, as a parameter. As may be seen, delay violation probability increases, while packet loss probability decreases as $L$ increases, which shows the tradeoff between delay violation and the packet loss probabilities.

Fig. 7,8,9 present the similar delay violation probabilities of data packets with respect to different parameters in a system with four types of traffics. These three numerical results show the same trend as in the system with two types of traffics.

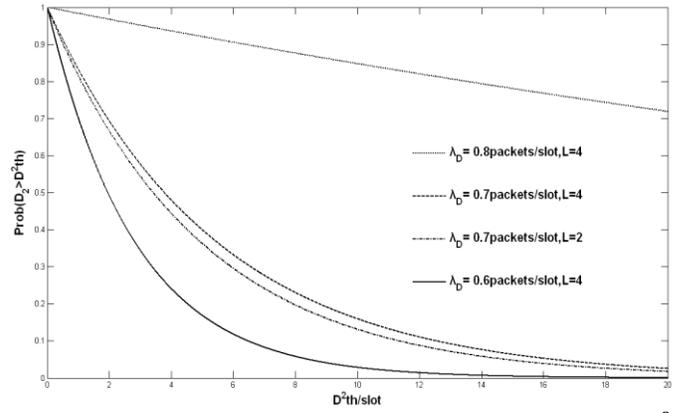
Fig. 4. Probability of queuing delay being greater than a threshold value $D_{th}^2$, for $\lambda_V = 0.1$ packets/slot, $p = 0.1$ wrt to different values of $\lambda_D$ and $L$

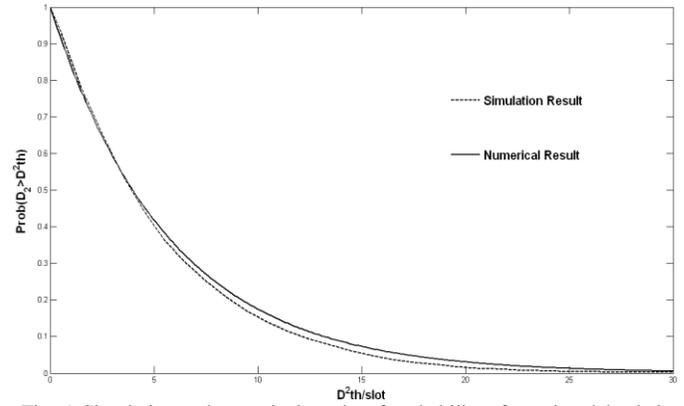
Fig. 5. Simulation and numerical results of probability of queuing delay being greater than a threshold value $D_{th}^2$, for $\lambda_D = 0.6$ packets/slot, $\lambda_V = 0.2$ packets/slot, $p = 0.1$, $L = 4$

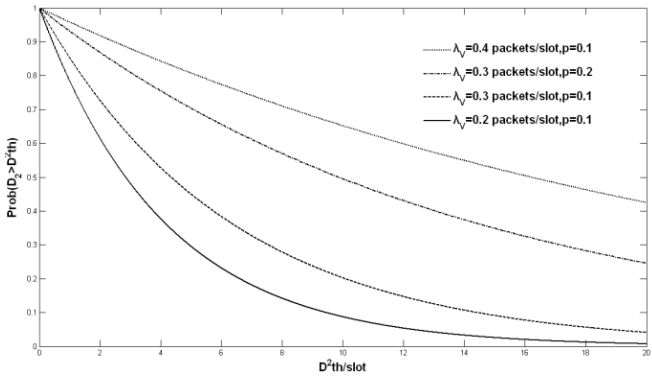
Fig. 3. Probability of queuing delay being greater than a threshold value $D_{th}^2$, for $\lambda_D = 0.5$ packets/slot, $L = 4$, wrt different values of $\lambda_v$ and $p$

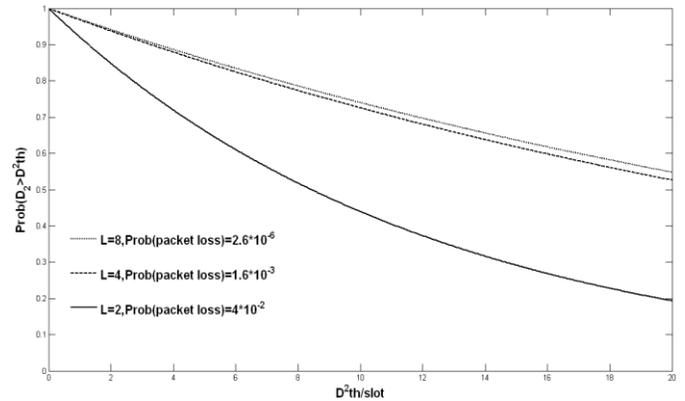
Fig. 6. Probability of queuing delay being greater than a threshold value $D_{th}^2$, for $\lambda_D = 0.7$ packets/slot, $\lambda_V = 0.1$ packets/slot, $p = 0.2$ wrt different values of $L$

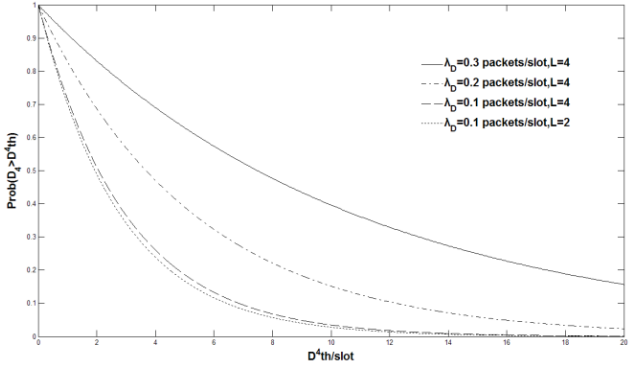

Fig. 7. Probability of queuing delay being greater than a threshold value $D_{th}^4$, for $\lambda_V = 0.4$packets/slot, $\lambda_A^1 = 0.1$packets/slot, $\lambda_A^2 = 0.2$packets/slot, $\pi_A^1 = 0.4, \pi_A^2 = 0.6, \lambda_B^1 = 0.3$packets/slot, $\lambda_B^2 = 0.2$packets/slot, $\pi_B^1 = 0.7, \pi_B^2 = 0.3, p = 0.1$ wrt different values of $\lambda_D$ and $L$

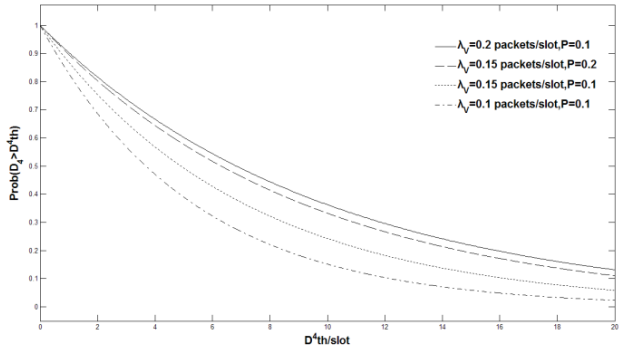

Fig. 8. Probability of queuing delay being greater than a threshold value $D_{th}^4$, for $\lambda_A^1 = 0.1$packets/slot, $\lambda_A^2 = 0.2$packets/slot, $\pi_A^1 = 0.4, \pi_A^2 = 0.6, \lambda_B^1 = 0.3$packets/slot, $\lambda_B^2 = 0.2$packets/slot, $\pi_B^1 = 0.7, \pi_B^2 = 0.3, \lambda_D = 0.4$packets/slot, $L = 4$, wrt different values of $\lambda_V$ and $p$

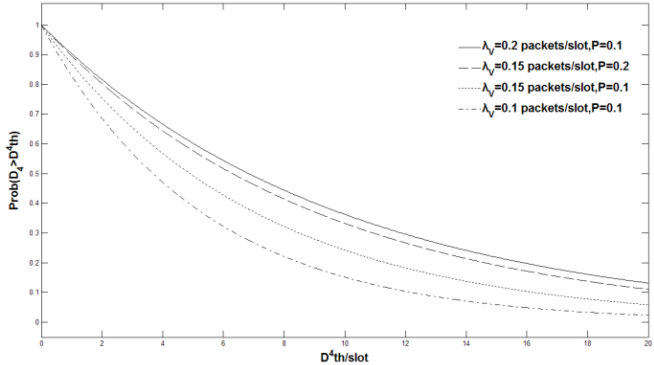

Fig. 9. Probability of queuing delay being greater than a threshold value $D_{th}^4$, for $\lambda_V = 0.2$packets/slot, $\lambda_B^1 = 0.3 packets/slot, \lambda_B^2 = 0.2 packets/slot$, $\pi_B^1 = 0.7, \pi_B^2 = 0.3, \lambda_D = 0.2 packets/slot, L = 4, p = 0.1$ wrt different values of $\lambda_A^1, \lambda_A^2, \pi_A^1$ and $\pi_A^2$

Next, we present numerical and simulation results about the analysis of the antenna placement part in the paper as well as simulation results. The numerical results for the optimal location of the antennas have been obtained through the application of the Robbins-Monro algorithm [15], the steps of which are given below,

a. Initialize the antennas location vector $L$ and let this vector to be $L(1)$
b. Randomly generate a user location vector $(\ell, \theta, 0)$ according to their uniform distributions in the cells.
c. Update the optimum locations of the antennas by applying the equation below

$$L(n+1) = L(n) + c_n \left(\frac{\partial \mathbb{P}(outage)}{\partial L}\right)\bigg| L = L(n)$$

d. Let $n=n+1$ and iteratively run steps b, c, d until $\overline{L(n)}$ converges, where $\overline{L(n)} = \frac{1}{n-1}\sum_{i=1}^{n-1} L(i)$. $\overline{L(n)}$ converges to the optimum antenna location vector $L'$.

In this implementation of the Robbins-Monro algorithm, we chose the sequence of positive $c_n$ as $c_n = 15n^{-0.75}$, which satisfies the Polyak and Juditsky constraints in [16].

We compared delay performance of the proposed algorithm with the case algorithm, where bandwidth slots are distributed between the data and voice traffics with different ratios in the case algorithm. Considering a TDMA system, the entire slot has 20 sub-slots and each sub-slot has a frame for potential voice or data transmission. The entire slot has both data and voice frames. The different ratios (Figures 12, 13) represent number of frames of data within the entire slot. Let $r$ represent this ratio. When there is no data or voice arrival, the TDMA system will transmit empty frames. Using the case algorithm, in spite of sensitivity to the data delay (as is shown in Figure 12), voice delay increases significantly (as is shown in Figure 13). However, the proposed algorithm has the lowest voice delay and doesn't have much data delay sensitivity.

Also, to reduce the amount of computations for the execution of the algorithm, we assumed that the optimal locations of the antennas will be a circle centered at the cell center and the antennas will be spaced evenly on the circle. The reason for this assumption is that the cells are symmetric in every aspect. The number of antennas is assumed to be $M=4$ and the cluster size to be $F=7$, $R=1$ bits/sec/Hz. In figures 5-10 wireless channel parameters are set as follows: $\hbar = 0.05$ and path loss exponent $2\lambda = 4$. The application of the Robbins-Monro procedure gives the optimum locations of antennas as $(L_m, \Theta_m, \hbar) = (0.58, \frac{\pi}{2}m - \frac{\pi}{2}, 0.05)$, for $m=1,…4$, which results in the minimum expected value of outage probability of, $E(\mathbb{P}) = 0.0087$, see Table 1.

In fig 11-13 wireless channel parameters are set as follows: $\hbar = 0.05$ and path loss exponent $2\lambda = 2$. In this case, Robbins-Monro procedure gives the optimum locations of antennas as $(L_m, \Theta_m, \hbar) = (0.416, \frac{\pi}{2}m - \frac{\pi}{2}, 0.05)$, for $m=1,…4$, which results in the minimum expected value of outage probability of, $E(\mathbb{P}) = 0.1081$, see Table 2.

In Fig. 10, we plotted simulation results for the expected value of the system outage probability as a function for the optimal radius of the antenna circle for different values of neighbor users' activation probability. From the figure it may be seen that there is an optimal radius of the antenna circle, 0.42, which results (for α = 1 ) in the minimum expected value of outage probability of E(ℙ) = 0.106 Comparing to the case that all the antennas are at the center, the optimal location provides almost 33% improvements in system's average outage probability. Simulation and numerical results show a very good match.

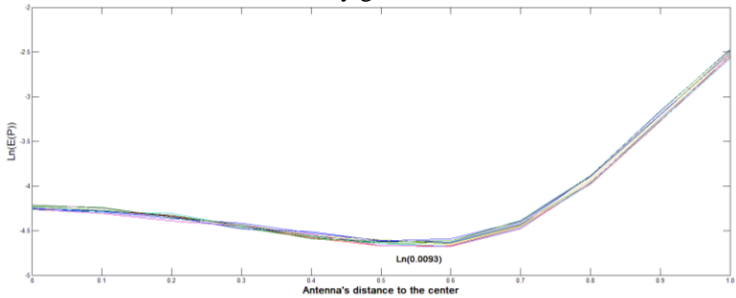

Fig. 10: Simulation result for system's average outage probability $E(\mathbb{P})$ as a function of the first antenna's location. (different curves represents for different value of $\Theta_1 = (0\frac{\pi}{18}, \frac{2\pi}{18}, \ldots, \frac{8\pi}{18})$ )

| n | $\overline{L_1}$ | E(p) | n | $\overline{L_1}$ | E(p) |
|---|---|---|---|---|---|
| 1 | 0.00 | 0.013 | 35 | 0.55 | 0.0095 |
| 2 | 0.90 | 0.091 | 36 | 0.55 | 0.0094 |
| 3 | 0.63 | 0.011 | 37 | 0.55 | 0.0094 |
| 4 | 0.54 | 0.009 | 38 | 0.55 | 0.0094 |
| 5 | 0.50 | 0.0099 | 39 | 0.55 | 0.0094 |
| 6 | 0.47 | 0.010 | 40 | 0.55 | 0.0094 |
| 7 | 0.45 | 0.010 | 41 | 0.55 | 0.0094 |
| 8 | 0.46 | 0.010 | 42 | 0.56 | 0.0094 |
| 9 | 0.48 | 0.010 | 43 | 0.56 | 0.0094 |
| 10 | 0.49 | 0.010 | 44 | 0.56 | 0.0093 |
| 11 | 0.50 | 0.0099 | 45 | 0.56 | 0.0093 |
| 12 | 0.50 | 0.0099 | 46 | 0.56 | 0.0093 |
| 13 | 0.50 | 0.0099 | 47 | 0.56 | 0.0093 |
| 14 | 0.51 | 0.0098 | 48 | 0.56 | 0.0093 |
| 15 | 0.51 | 0.0097 | 49 | 0.57 | 0.0093 |
| 16 | 0.51 | 0.0097 | 50 | 0.57 | 0.0093 |
| 17 | 0.51 | 0.0096 | 51 | 0.57 | 0.0093 |
| 18 | 0.51 | 0.0096 | 52 | 0.57 | 0.0092 |
| 19 | 0.52 | 0.0096 | 53 | 0.57 | 0.0092 |
| 20 | 0.52 | 0.0096 | 54 | 0.57 | 0.0092 |
| 21 | 0.52 | 0.0095 | 55 | 0.57 | 0.0092 |
| 22 | 0.52 | 0.0095 | 56 | 0.57 | 0.0092 |
| 23 | 0.53 | 0.0095 | 57 | 0.58 | 0.0091 |
| 24 | 0.53 | 0.0095 | 58 | 0.58 | 0.0091 |
| 25 | 0.53 | 0.0095 | 59 | 0.58 | 0.0091 |
| 26 | 0.54 | 0.0095 | 60 | 0.58 | 0.0090 |
| 27 | 0.54 | 0.0095 | 61 | 0.58 | 0.0090 |
| 28 | 0.54 | 0.0095 | 62 | 0.58 | 0.0090 |
| 29 | 0.54 | 0.0095 | 63 | 0.58 | 0.0090 |
| 30 | 0.54 | 0.0095 | 64 | 0.58 | 0.0090 |
| 31 | 0.54 | 0.0095 | 65 | 0.58 | 0.0090 |
| 32 | 0.54 | 0.0095 | 66 | 0.58 | 0.0090 |
| 33 | 0.54 | 0.0095 | 67 | 0.58 | 0.0090 |
| 34 | 0.54 | 0.0095 | 68 | 0.58 | 0.0090 |

Table 1. Robbins-Monro outcome of the optimum location of the first antenna as. ( here we only show the optimum antenna radius $\overline{L_1(n)}$ since angles $\overline{\Theta_1(n)}$ changes slightly around 0 as *n* increases.)

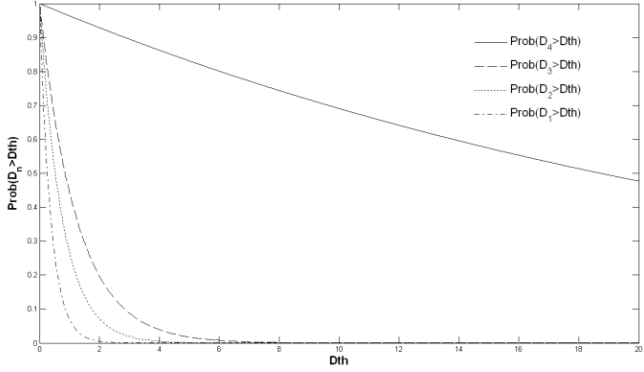

Fig. 11: Probability of queuing delay being greater than a threshold value $D_{th}$, for $\lambda_V = 0.2 \text{packets/slot}, \lambda_A^1 = 0.2 \text{packets/slot}, \lambda_A^2 = 0.3 \text{packets/slot}, \pi_A^1 = 0.8, \pi_A^2 = 0.2, \lambda_B^1 = 0.3 \text{packets/slot}, \lambda_B^2 = 0.2 \text{packets/slot}, \pi_B^1 = 0.7, \pi_B^2 = 0.3, \lambda_D = 0.2 \text{packets/slot}, L = 4, p = 0.1$

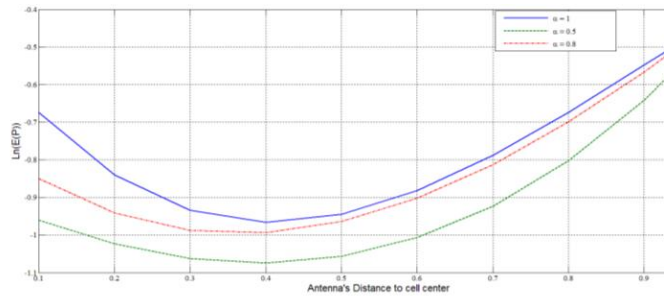

Fig. 12: Simulation result for system's average outage probability $E(\mathbb{P})$ as a function of the first antenna's location. (different curves represents

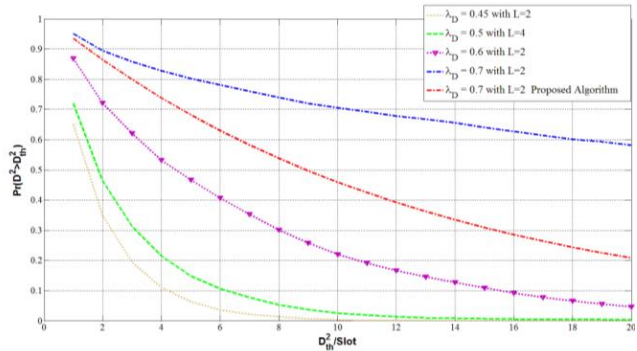

Fig. 13: Probability of data queuing delay being greater than a threshold value $D_{th}^2$, for $\lambda_V = 0.1$ packets/slot, $p = 0.1$ and $r = 0.85$

| $n$ | $\overline{L_1}$ | E(p) | $n$ | $\overline{L_1}$ | E(p) | $n$ | $\overline{L_1}$ | E(p) |
|---|---|---|---|---|---|---|---|---|
| 1 | 0.9 | 0.293 | 18 | 0.37 | 0.114 | 35 | 0.40 | 0.11 |
| 2 | 0.89 | 0.293 | 19 | 0.37 | 0.114 | 36 | 0.40 | 0.11 |
| 3 | 0.71 | 0.173 | 20 | 0.37 | 0.113 | 37 | 0.40 | 0.11 |
| 4 | 0.54 | 0.119 | 21 | 0.37 | 0.113 | 38 | 0.40 | 0.11 |
| 5 | 0.44 | 0.109 | 22 | 0.38 | 0.113 | 39 | 0.40 | 0.11 |
| 6 | 0.35 | 0.115 | 23 | 0.38 | 0.113 | 40 | 0.40 | 0.109 |
| 7 | 0.34 | 0.115 | 24 | 0.38 | 0.113 | 41 | 0.40 | 0.109 |
| 8 | 0.35 | 0.115 | 25 | 0.38 | 0.113 | 42 | 0.40 | 0.109 |
| 9 | 0.35 | 0.115 | 26 | 0.38 | 0.113 | 43 | 0.40 | 0.109 |
| 10 | 0.35 | 0.115 | 27 | 0.38 | 0.113 | 44 | 0.41 | 0.108 |
| 11 | 0.36 | 0.114 | 28 | 0.39 | 0.113 | 45 | 0.41 | 0.108 |
| 12 | 0.36 | 0.114 | 29 | 0.39 | 0.113 | 46 | 0.41 | 0.108 |
| 13 | 0.36 | 0.114 | 30 | 0.39 | 0.113 | 47 | 0.41 | 0.108 |
| 14 | 0.36 | 0.114 | 31 | 0.39 | 0.113 | 48 | 0.41 | 0.108 |
| 15 | 0.36 | 0.114 | 32 | 0.39 | 0.112 | 49 | 0.41 | 0.108 |
| 16 | 0.37 | 0.114 | 33 | 0.39 | 0.112 | 50 | 0.41 | 0.108 |
| 17 | 0.37 | 0.114 | 34 | 0.39 | 0.112 | 51 | 0.41 | 0.108 |

Table 2. Robbins-Monro outcome of the optimum location of the first antenna as. ( here we only show the optimum antenna radius $\overline{L_1(n)}$ since angles $\overline{\theta_1(n)}$ changes slightly around 0 as *n* increases.)

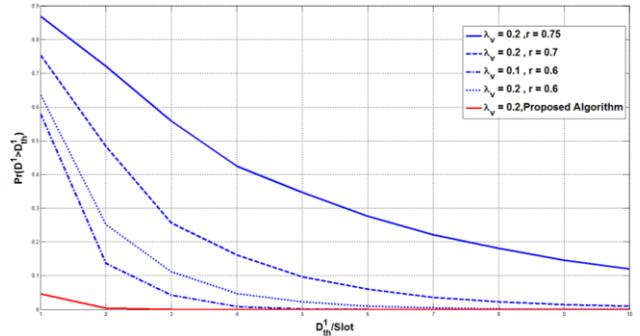

Fig. 14: Probability of voice queuing delay being greater than a threshold value $D_{th}^1$, for $\lambda_D = 0.5$ packets/slot, $p = 0.1$

## VIII. CONCLUSION

Presently, distributed broadband wireless communication (BWC) system is a very promising wireless technology due to its low cost and high performance. This paper considers uplink communications with multiple levels of priority traffic having any renewal arrival and departure processes. We develop an analysis that determines packet delay violation probability for each priority level as a function of the outage probability of the distributed antenna system through the application of results from the renewal theory. Then, we determine the optimal locations of the antennas that minimize the antenna outage probability. We also present simulation results that show the accuracy of the analysis. The results are very easy to use and they will be useful in the design of BWC systems. Finally, the packet delay violation probability result may also be applicable in any other communication systems with multiple types of priority traffic.